\documentstyle[aps,epsf]{revtex}

\title{
       The Gross-Neveu Model and the Supersymmetric
       and Non--Supersymmetric Nambu--Jona--Lasinio 
       Model in a Magnetic Field
} 
\author{
        V.Elias$^{1}$, 
        D.G.C.McKeon$^{1}$, 
        V.A.Miransky$^{1,2}$,  
   and  I.A.Shovkovy$^{1,2}$
}
\address{
         $^{1}$Department of Applied Mathematics,\\
               University of Western Ontario,\\
               London, ON N6A 5B7, Canada\\
         $^{2}$Bogolyubov Institute for 
                      Theoretical Physics,\\
               252143 Kiev, Ukraine
} 

\date{December 15, 1996}

\draft

\begin{document}

\maketitle

\begin{abstract}
The infrared dynamics in the ($3+1$)--dimensional 
supersymmetric and non--supersymmetric Nambu--Jona--Lasinio 
model in a constant magnetic field is studied.  
While at strong coupling the dynamics in these two models 
is essentially different, it is shown that the models become 
equivalent at weak coupling. In particular, at weak coupling,
as the strength of the magnetic field goes to infinity, both 
the supersymmetric and non--supersymmetric Nambu--Jona--Lasinio 
models are reduced to a continuum set of independent 
(1+1)--dimensional Gross--Neveu models, labeled by the 
coordinates in the plane perpendicular to the magnetic field. 
The relevance of these results for cosmological models 
based on supersymmetric dynamics is pointed out. 
\end{abstract}

\pacs{11.30.Rd, 11.30.Qc, 12.60.Jv, 12.20.Ds}

\section{Introduction.}

It has been recently shown \cite{1,2,3,4} that a constant magnetic
field in $2+1$ and $3+1$ dimensions is a strong catalyst of dynamical
chiral symmetry breaking, leading to the generation of a fermion
dynamical mass even at the weakest attractive interaction between
fermions.

The essence of the effect is the dimensional reduction $D\to 
D-2$ (i.e. $2+1 \to 0+1$ and $3+1 \to 1+1$ ) in the infrared 
dynamics of the fermion pairing in a magnetic field. The 
physical reason of this reduction is the fact that the motion of 
charged particles is restricted in those directions that are 
perpendicular to the magnetic field.  This is in turn connected 
with the point that, at weak coupling between fermions, the 
fermion pairing, leading to the chiral condensate, is mostly 
provided by fermions from the lowest Landau level (LLL) whose 
dynamics are ($D-2$)--dimensional.

In this paper, we shall further clarify the effect of the 
dynamical reduction, studying in detail the infrared dynamics in 
the ($3+1$)--dimensional Nambu--Jona--Lasinio (NJL) model in a 
magnetic field.  We shall consider both the ordinary 
(non--supersymmetric) and supersymmetric versions of the NJL 
model. In particular, we shall show that in the ``continuum" 
limit, when both the strength of the magnetic field and the 
ultraviolet cutoff go to infinity, both the non--supersymmetric 
and supersymmetric (weakly coupling) NJL models with $N_c$ 
colors are reduced to a continuum set of independent 
(1+1)--dimensional Gross-Neveu (GN) models \cite{5}, labeled by 
coordinates ${\bf x}_{\perp}$ in the plane perpendicular to the 
magnetic field ${\bf B}$.  The number of colors in the GN models 
is $\tilde{N}_c=(\pi/2C)N_c$, where $C$ is $C=\Lambda^2/|eB|$ in 
the ``continuum" limit (here $\Lambda$ is the ultraviolet 
cutoff).  As will be shown in Sec.3, the factor $\pi/2C$ is 
proportional to a (local) magnetic flux attached to each point 
in the $x_{\perp}$-plane.

On the other hand, at strong coupling, the dynamics in the 
supersymmetric and non--supersymmtric NJL models are very 
different.

Recall that there is no spontaneous chiral symmetry breaking in 
the supersymmetric NJL model \cite{6}. An external magnetic 
field changes the situation dramatically: chiral symmetry 
breaking occurs for any value of the coupling constant in this 
model. This agrees with the general conclusion of 
Refs.~\cite{2,3,4} that the effect of the catalysis of chiral 
symmetry breaking by a magnetic field in $3+1$ dimensions is a 
universal, model--independent effect.

As was already shown in Ref.~\cite{4}, the dimensional reduction 
$3+1 \to 1+1$ in the dynamics of the fermion pairing in a 
(finite) magnetic field is consistent with spontaneous symmetry 
breaking. Recall that, due to the Mermin--Wagner--Coleman (MWC) 
theorem \cite{7}, there cannot be spontaneous breakdown of 
continuous symmetries at $D=1+1$.    The MWC theorem is based on 
the fact that gapless Nambu--Goldstone (NG) bosons cannot exist 
in $1+1$ dimensions. However, in a magnetic field, the reduction 
$3+1 \to 1+1$ takes place (in the infrared region) only for 
propagators of charged particles: it reflects the fact that the 
motion of charged particles is restricted in the directions 
perpendicular to the magnetic field. On the other hand, NG 
bosons connected with spontaneous chiral symmetry breaking are 
neutral and therefore their propagators have 
($3+1$)--dimensional form \cite{4}\footnote{The Lorentz 
invariance is broken by a magnetic field in this problem. By the 
($3+1$)--dimensional form, we understand that the denominator of 
the propagators depends on energy and all the components of the 
center--of--mass momentum, i.e.  ${\cal 
D}(P)\sim(P_0^2-C_{\parallel}P_{\parallel}^2- C_{\perp}{\bf 
P}_{\perp}^2)^{-1}$ with $C_{\parallel},C_{\perp}\neq 0$.}.  
This in turn implies that the effect of spontaneous chiral 
symmetry breaking does not contradict the MWC theorem.

The interplay between the GN model and the NJL model in a 
magnetic field established in this paper further clarifies this 
issue. As was mentioned above, the NJL model is reduced to a set 
of the GN models only as the strength of the magnetic field goes 
to infinity. As we shall show in Sec.4, at finite $|eB|$, the 
dynamics of the NJL model in a magnetic field are in a sense 
similar to the dynamics in the ($2+\epsilon$)--dimensional GN 
model: the magnetic length $l=|eB|^{-1/2}$ plays here the role 
of the (physical) $\epsilon$--parameter which is an infrared 
regulator.

As was already pointed out in Refs.\cite{1,2,3,4}, there may be 
interesting applications of this effect in cosmology as well as 
in particle and condensed matter physics. The results of the 
present paper may be particularly relevant for cosmological 
scenarios based on supersymmetric dynamics \cite{9}.

The paper is organized as follows. In Section 2 we, for 
completeness, derive the effective action in the GN model. In 
Sections 3 and 4 we establish the connection between the GN 
model and the NJL model in a magnetic field as $|eB|\to\infty$.  
In Section 5 we consider the dynamics of the supersymmetric NJL 
model in a magnetic field. In Section 6 we summarize the main 
results of the paper and discuss possible applications of these 
results and as well as the possibility of their extension to 
inhomogeneous magnetic field configurations.  In the Appendix 
some useful formulas and relations are derived.

\section{Effective Action in the Gross--Neveu Model.}

In this section, for completeness, we shall derive the effective 
action for the GN model. The Lagrangian density of the GN model 
is:
\begin{equation}
{\cal L}_{GN} = \frac{1}{2} 
\left[\bar{\Psi}, (i\gamma^\mu \partial_\mu)\Psi\right] +
\frac{\tilde{G}}{2} 
\left[ (\bar{\Psi}\Psi)^2+(\bar{\Psi}i\gamma^5\Psi)^2 \right]
\label{eq1}
\end{equation}
where $\mu=0,1$ and the fermion field carries an additional 
``color" index $\tilde{\alpha}=1,2,\dots,\tilde{N}_c$ (for 
simplicity, we consider the case of the chiral $U_L(1)\times 
U_R(1)$ symmetry). The theory is equivalent to the theory with 
the Lagrangian density
\begin{equation}
{\cal L}^{\prime}_{GN}=\frac{1}{2} 
\left[\bar{\Psi}, (i\gamma^\mu  \partial_\mu) \Psi\right] 
- \bar{\Psi}(\sigma+i\gamma^5\pi)\Psi
- \frac{1}{2\tilde{G}} \left(\sigma^2+\pi^2\right).
\label{eq2}
\end{equation}
The Euler--Lagrange equations for the auxiliary fields $\sigma$ and
$\pi$ take the form of constraints:
\begin{equation} 
\sigma=-\tilde{G}\bar{\Psi}\Psi, \qquad 
\pi=-\tilde{G}\bar{\Psi}i\gamma^5\Psi,
\label{eq3}
\end{equation}
and the Lagrangian density (\ref{eq2}) reproduces Eq.(\ref{eq1}) upon
application of the constraints (\ref{eq3}). The effective action for
the composite fields $\sigma$ and $\pi$ can be obtained by
integrating over fermions in the path integral. It is given by the
standard relation: 
\begin{eqnarray}
\Gamma_{GN}(\sigma,\pi) &=& \tilde{\Gamma}_{GN}(\sigma,\pi)
-\frac{1}{2\tilde{G}}\int d^2x(\sigma^2+\pi^2), 
\label{eq4} \\
\tilde{\Gamma}_{GN}(\sigma,\pi) &=&- i Tr Ln \left[i\gamma^\mu
\partial_\mu - (\sigma+i\gamma^5\pi)\right].
\label{eq5}
\end{eqnarray}
The low energy quantum dynamics are described by the path 
integral (with the integrand $\exp(i\Gamma_{GN})$) over the 
composite fields $\sigma$ and $\pi$. As $\tilde{N}_c\to\infty$, 
the path integral is dominated by the stationary points of the 
action:  
$\delta\Gamma_{GN}/\delta\sigma=\delta\Gamma_{GN}/\delta\pi=0$. 
We will analyze the dynamics by using the expansion of the 
action $\Gamma_{GN}$ in powers of derivatives of the composite 
fields.

We begin the calculation of $\Gamma_{GN}$ by calculating the 
effective potential $V_{GN}$. Since $V_{GN}$ depends
only on the $U_L(1)\times U_R(1)$--invariant $\rho^2=\sigma^2+\pi^2$, 
it is sufficient to consider a configuration with $\pi=0$ and $\sigma$ 
independent of $x$. Then we find from Eqs.~(\ref{eq4}) and (\ref{eq5}):
\begin{eqnarray}
V_{GN}(\rho)&=&\frac{\rho^2}{2\tilde{G}} 
             - \tilde{N}_c\int \frac{d^2k}{(2\pi)^2} 
               \ln\left(\frac{k^2+\rho^2}{k^2}\right)=
\nonumber\\
            &=&\frac{\rho^2}{2\tilde{G}} 
             - \frac{\tilde{N}_c\rho^2}{4\pi}
               \left[\ln\frac{\Lambda^2}{\rho^2}+1\right],
\label{eq6}
\end{eqnarray}
where the integration is done in Euclidean region ($\Lambda$ is an
ultraviolet cutoff). As is known, in the GN model, the equation of
motion $dV_{GN}/d\rho=0$ has a nontrivial solution
$\rho =\bar{\sigma}\equiv m_{dyn}$ for any value of the coupling
constant $\tilde{G}$. Then the potential $V_{GN}$ can be rewritten as 
\begin{equation}
V_{GN}(\rho)=\frac{\tilde{N}_c\rho^2}{4\pi}
\left[\ln\frac{\rho^2}{m_{dyn}^2}-1\right],
\label{eq7}
\end{equation}
where
\begin{equation}
m_{dyn}^2=\Lambda^2\exp
\left(-\frac{2\pi}{\tilde{N}_c\tilde{G}}\right).
\label{eq8}
\end{equation}
Due to the MWC theorem \cite{7}, there cannot be spontaneous
breakdown of continuous symmetries at $D=1+1$. The parameter $m_{dyn}$
is an order parameter of chiral symmetry breaking only in leading
order in $1/\tilde{N}_c$ (this reflects the point that the MWC
theorem is not applicable to systems with $\tilde{N}_c\to \infty$
\cite{8}). In the exact GN solution, spontaneous chiral symmetry
breaking is washed out by interactions (strong fluctuations) of
would--be NG bosons $\pi$ (i.e. after integration over $\pi$ and 
$\sigma$ in the path integral). The exact solution in 
this model presumably corresponds to the realization of the
Berezinsky--Kosterlitz--Thouless (BKT) phase: though chiral
symmetry is not broken in this phase, 
the parameter $m_{dyn}$ still defines the
fermion mass, and the would--be NG boson $\pi$ transforms into a BKT
gapless excitation \cite{8}.

Let us now turn to calculating the kinetic term in $\Gamma_{GN}$. The
chiral $U_L(1)\times U_R(1)$ symmetry implies that the general form
of the kinetic term is
\begin{equation}
{\cal L}^{(k)}_{GN} = \frac{f_1^{\mu\nu}}{2} (\partial_\mu\rho_j\partial_\nu \rho_j)
+ \frac{f^{\mu\nu}_2}{\rho^2}(\rho_j\partial_\mu\rho_j)(\rho_i
\partial_\nu\rho_i)
\label{eq9}
\end{equation}
where $\mbox{\boldmath$\rho$}=(\sigma,\pi)$ and $f^{\mu\nu}_1$, 
$f^{\mu\nu}_2$ are functions of $\rho^2$. To find the functions 
$f^{\mu\nu}_1$ and $f^{\mu\nu}_2$, one can use different methods. We
utilize the same method as in Ref.~\cite{4} (see Appendix A in that
paper). The result is:
\begin{eqnarray}
f^{\mu\nu}_1(\rho^2)&=&-\frac{i}{2}
\int\frac{d^2k}{(2\pi)^2} tr\left[
S(k)i\gamma_5\frac{\partial^2S(k)}{\partial k_{\mu}\partial k_{\nu}}
i\gamma_5\right],
\label{eq10} \\
f^{\mu\nu}_2(\rho^2)&=&-\frac{i}{4}
\int\frac{d^2k}{(2\pi)^2} tr\left[
S(k)\frac{\partial^2S(k)}{\partial k_{\mu}\partial k_{\nu}}-
S(k)i\gamma_5\frac{\partial^2S(k)}{\partial k_{\mu}\partial k_{\nu}}
i\gamma_5\right],
\label{eq11}
\end{eqnarray}
with  $S(k)=i(k^{\mu}\gamma_{\mu}+\rho)/(k^2-\rho^2)$. The
explicit form of these functions is:
\begin{equation}
f^{\mu\nu}_1=g^{\mu\nu}\frac{\tilde{N}_c}{4\pi\rho^2},\qquad
f^{\mu\nu}_2=-g^{\mu\nu}\frac{\tilde{N}_c}{12\pi\rho^2}
\label{eq12}
\end{equation}

\section{The Interplay between the GN Model and the 
NJL Model in a Magnetic Field}

In this section, we compare the effective actions in the GN model and
in the NJL model in a magnetic field, and we establish a rather
interesting connection between these two models.

The analog of the Lagrangian density (\ref{eq2}) in the NJL model in
a magnetic field is 
\begin{equation}
{\cal L}^{\prime}=\frac{1}{2} 
    \left[\bar{\Psi}, (i\gamma^\mu  D_\mu) \Psi\right] 
- \bar{\Psi}(\sigma+i\gamma^5\pi)\Psi
- \frac{1}{2G} \left(\sigma^2+\pi^2\right)
\label{eq2njl}
\end{equation}
where $D_\mu=\partial_\mu - ie A_\mu^{ext}$, 
$A_\mu^{ext}=Bx^2\delta_\mu^3$
(the magnetic field is in $+x_1$ direction).

In leading order in $1/N_c$, the effective action 
in the NJL model in a magnetic field is derived
in Refs.~\cite{2,4}. The effective potential and the kinetic term are
($\mbox{\boldmath$\rho$}=(\sigma,\pi)$):
\begin{eqnarray}
V(\rho)&=&\frac{\rho^2}{2G}+\frac{N_c}{8\pi^2} 
       \Bigg[\frac{\Lambda^4}{2}
     + \frac{1}{3l^4}\ln(\Lambda l)^2 
     + \frac{1-\gamma-\ln2}{3l^4}-(\rho\Lambda)^2
     + \frac{\rho^4}{2}\ln(\Lambda l)^2  
 \nonumber\\
 &+&\frac{\rho^4}{2}(1-\gamma-\ln2)
     + \frac{\rho^2}{l^2}\ln\frac{(\rho l)^2}{2}
     - \frac{4}{l^4}\zeta^{\prime} 
       (-1,\frac{(\rho l)^2}{2}+1)\Bigg]
     + O\left(\frac{1}{\Lambda}\right),
\label{eq13}
\end{eqnarray}
\begin{equation}
{\cal L}^{(k)} = \frac{f_1^{\mu\nu}}{2} 
         (\partial_\mu\rho_j\partial_\nu \rho_j)
       + \frac{f^{\mu\nu}_2}{\rho^2}
         (\rho_j\partial_\mu\rho_j)(\rho_i \partial_\nu\rho_i)
\label{eq14}
\end{equation}
with $f^{\mu\nu}_1$ and 
     $f^{\mu\nu}_2$ 
being diagonal tensors:
\begin{eqnarray}
f^{00}_1 = - f^{11}_1&=&\frac{N_c}{8\pi^2}
           \left[\ln\frac{(\Lambda l)^2}{2}
         - \psi\left(\frac{(\rho l)^2}{2}+1\right)
         + \frac{1}{(\rho l)^2} - \gamma
         + \frac{1}{3} \right], 
         \nonumber\\
f^{22}_1 = f^{33}_1&=& - \frac{N_c}{8\pi^2}
           \left[\ln\frac{\Lambda^2}{\rho^2}
         - \gamma+\frac{1}{3}\right], 
         \nonumber\\
f^{00}_2 = - f^{11}_2&=&-\frac{N_c}{24\pi^2}
           \left[\frac{(\rho l)^2}{2}
           \zeta\left(2,\frac{(\rho l)^2}{2}+1\right)
         + \frac{1}{(\rho l)^2}\right], 
         \label{eq15} \\
f_2^{22} = f^{33}_2&=& - \frac{N_c}{8\pi^2}
    \Bigg[(\rho l)^4 \psi \left(\frac{(\rho l)^2}{2}+1\right)
         - 2(\rho l)^2\ln\Gamma\left(\frac{(\rho l)^2}{2}\right)
         \nonumber\\
        &&\qquad - (\rho l)^2 \ln\frac{(\rho l)^2}{4\pi}
        - (\rho l)^4 - (\rho l)^2+1\Bigg].
\nonumber
\end{eqnarray}
Here $G$ is the NJL coupling constant, $N_c$  is the number of
colors, $\zeta(\nu,x)$ is the generalized Riemann zeta function,
$\zeta^{\prime}(\nu,x)=\partial\zeta(\nu,x)/\partial\nu$,  
$\gamma\approx0.577$ is the Euler constant,
$\psi(x)=d\left(\ln\Gamma(x)\right)/dx$,
and $l\equiv|eB|^{-1/2}$ is the magnetic length.
The gap equation 
$dV/d\rho=0$ is\footnote{In this paper we consider the case of a
large ultraviolet cutoff: $\Lambda^2\gg\bar{\sigma}^2$, $|eB|$, 
where $\bar{\sigma}$ is a minimum of the potential $V$.}
\begin{equation}
\rho\Lambda^2  \left(\frac{1}{g}-1\right)=
        - \rho^3\ln\frac{(\Lambda l)^2}{2}+\gamma\rho^3
        + \frac{\rho}{l^2}\ln\frac{(\rho l)^2}{4\pi}
        + \frac{2\rho}{l^2}\ln\Gamma\left(\frac{(\rho l)^2}{2}\right)
        + O\left(\frac{1}{\Lambda}\right),
\label{eq16}
\end{equation}
where the dimensionless coupling constant $g=N_cG\Lambda^2/4\pi^2$.
In the derivation of this equation, we used the relations \cite{10}:
\begin{eqnarray}
\frac{\partial}{\partial x}\zeta(\nu,x)=-\nu\zeta(\nu+1,x),&& 
                            \label{eq17} \\
\left.\frac{\partial}{\partial \nu}\zeta(\nu,x)\right|_{\nu=0}=
              \ln\Gamma(x)-\frac{1}{2}\ln2\pi,
           &\quad& \zeta(0,x)=\frac{1}{2}-x.
\label{eq18}
\end{eqnarray}
As $B\to0$ ($l\to\infty$), we recover the known gap equation in the NJL
model (for a review see Ref.\cite{11}):
\begin{equation}
\rho\Lambda^2\left(\frac{1}{g}-1\right)=
          - \rho^3\ln\frac{\Lambda^2}{\rho^2}.
\label{eq19}
\end{equation}
This equation admits a nontrivial solution only if $g$ is supercritical,
$g>g_c=1$ (as Eq.(\ref{eq2njl}) implies, a solution to the gap equation, 
$\rho=\bar{\sigma}$, coincides with the fermion dynamical mass, 
$\bar{\sigma}=m_{dyn}$). As was shown in Refs.~\cite{2,4}, 
at $B\neq0$, a non--trivial solution exists for all $g>0$.

Let us consider the case of small subcritical $g$, $g\ll g_c=1$, in
detail. A solution is seen to exist for this case if $\rho l$ is
small. Specifically, for $g\ll 1$, the left--hand side of 
Eq.(\ref{eq16}) is positive. Since the first term of the 
right--hand side in this equation is negative, we conclude 
that a non--trivial solution to this equation may exist only for
\begin{equation} 
\rho^2 \ln(\Lambda l)^2 \ll 
       \frac{1}{l^2} \ln\frac{1}{(\rho l)^2}
\label{eq20}
\end{equation}
($\Gamma(\rho^2 l^2/2)\approx 2/(\rho l)^2$). 
We then find the solution:
\begin{equation}
m^2_{dyn} \equiv \bar{\sigma}^2 = \frac{|eB|}{\pi}\exp
          \left(-\frac{4\pi^2(1-g)}{|eB| N_c G}\right) 
          = \frac{|eB|}{\pi}\exp
          \left(-\frac{(1-g)\Lambda^2}{g|eB|}\right).
\label{eq21}
\end{equation}
Actually, since Eq.(\ref{eq21}) implies that condition (\ref{eq20}) is
violated only if $(1-g) \alt |eB|/\Lambda^2$, the expression (\ref{eq21}) is
valid for all $g$ outside that (scaling) region near the critical
value $g_c=1$. Note that in the scaling region ($g\to g_c-0$) the
expression for $m_{dyn}^2$ is different \cite{2,4}:
\begin{equation}
m_{dyn}^2\simeq |eB|
            \frac{\ln\left[(\ln\Lambda^2l^2)/\pi \right]}
                 {\ln\Lambda^2l^2}.
\label{eq22}
\end{equation}
Let us compare relation (\ref{eq21}) with relation (\ref{eq8}) for the
dynamical mass in the GN model. The similarity between them is
evident: $|eB|$ and $|eB|G$ in Eq.(\ref{eq21}) play the role of an
ultraviolet cutoff and the dimensionless coupling constant $\tilde{G}$
in Eq.(\ref{eq8}). Let us discuss this connection and show that it is
intimately connected with the dimensional reduction $3+1\to 1+1$ in
the dynamics of the fermion pairing in a magnetic field.

Eq.(\ref{eq8}) implies that the GN model is asymptotically free, with
the bare coupling constant
$\tilde{G}=2\pi/\tilde{N}_c\ln(\Lambda^2/m_{dyn}^2)\to 0$ as
$\Lambda\to\infty$. Let us now consider the following limit in the
NJL model in a magnetic field: $|eB|\to\infty$, $\Lambda^2/|eB|=C\gg
1$. Then relation  (\ref{eq21}), which can be rewritten as
\begin{equation}
m^2_{dyn} = \frac{\Lambda^2}{C\pi}
            \exp\left(-\frac{C(1-g)}{g}\right),
\label{eq23}
\end{equation}
implies that the behavior of the bare coupling constant $g$ must be 
\begin{equation}
      g \simeq \frac{C}{\ln(\Lambda^2/C\pi m_{dyn}^2)} \to 0,
\label{eq24}
\end{equation}
in order to get a finite value for $m_{dyn}^2$ in this
limit. Thus in this ``continuum" limit, we recover the same behavior for 
the coupling $g$ in the NJL model as for the coupling constant
$\tilde{G}$ in the GN model.

Let us now compare the effective potentials in these two models. At
first glance, the expressions (\ref{eq6}) and (\ref{eq13}) for the
effective potentials in these models look very
different: the character of ultraviolet divergences in
1+1 and 3+1 dimensional theories 
is essentially different. However,
using Eqs.(\ref{eq17}) and (\ref{eq18}), the expression (\ref{eq13}) can be
rewritten, for small $\rho l$, as 
\begin{eqnarray}
V(\rho) &=& V(0) + \frac{N_c\rho^2}{8\pi^2l^2} 
    \Bigg[ (\Lambda l)^2 \left(\frac{1}{g}-1\right)
        - 1 + \ln(\pi\rho^2 l^2) +
    \nonumber\\
        && \qquad + \frac{(\rho l)^2}{2}\ln\frac{(\Lambda l)^2}{2}
        + O((\rho l)^4)\Bigg].
\label{eq25}
\end{eqnarray}
Then, expressing the coupling constant $g$ through $m_{dyn}$ from
Eq.(\ref{eq21}), we find that
\begin{eqnarray}
V(\rho) &=& V(0) + \frac{N_c|eB|}{8\pi^2}\rho^2
        \left[ \ln\frac{\rho^2}{m_{dyn}^2} - 1 + O((\rho l)^2) \right] .
\label{eq26}
\end{eqnarray}
Here we used the fact that, because of Eq.(\ref{eq20}), the ratio 
$(\rho l)^2$ is small near the minimum $\rho=m_{dyn}$.

The expressions (\ref{eq7}) and (\ref{eq26}) for the potentials in these two
models look now quite similar. There is however an additional factor
$|eB|/2\pi$ in the expression (\ref{eq26}). Moreover, the  field 
$\rho$, which depends on the two coordinates $x_0$ and
$x_1$ in the GN model, depends on the four coordinates 
$x_0$, $x_1$, $x_2$ and $x_3$ in the NJL model.

In order to clarify this point, let us turn to the analysis of the
kinetic term (\ref{eq14}) in the effective action of the NJL model
in a magnetic field.

Because of the expression (\ref{eq21}) for $m_{dyn}$ at small $g$, the
term $1/(\rho l)^2$ dominates in the functions $f_{1}^{00}=-f_{1}^{11}$ 
and $f_{2}^{00}=-f_{2}^{11}$, around the minimum $\rho=m_{dyn}$:
\begin{eqnarray}
f^{00}_1=-f^{11}_1 \simeq
        \frac{N_c}{8\pi^2}\frac{1}{\rho^2l^2},\quad
f^{00}_2=-f^{11}_2 \simeq
        -\frac{N_c}{24\pi^2}\frac{1}{\rho^2l^2}.
\label{eq27}
\end{eqnarray}
Up to the additional factor $|eB|/2\pi$, these functions coincide
with those in (\ref{eq12}) in the GN model.
On the other hand, the functions $f^{22}_1=f^{33}_1$ and
$f^{22}_2=f^{33}_2$, connected with derivatives with respect to the
transverse coordinates, are strongly supressed, as compared to the
functions (\ref{eq27}), and the ratios of the functions
$f^{22}_1=f^{33}_1$ and $f^{22}_2=f^{33}_2$ to those in
(\ref{eq27}) go rapidly (as $m^{2}_{dyn}/\Lambda^2$) to zero as 
$|eB|\to \infty$, $\Lambda^2/|eB|=C$.                                 

As a result, the coordinates $x_2$ and $x_3$ become redundant
variables in this limit: there are no transitions of field quanta
between different points in the $x_{2}x_{3}$--plane. Therefore
the model degenerates into a set of independent $(1+1)$--dimensional
models, labeled by ${\bf x}_{\perp}=(x_2,x_3)$ coordinates.
Let us show that all these models coincide with a $(1+1)$--dimensional
GN model with the number of colors
$\tilde{N}_c=(\pi/2C)N_c$, where the factor $C=\Lambda^2/|eB|$ was
introduced above, in the definition of the ``continuum" limit
($|eB|\to\infty$, $\Lambda^2/|eB|=C$).

Let us put the NJL model 
on a lattice with $a$ the lattice 
spacing of the discretized space-time (in Euclidean region).
Then its effective action can be
written as
\begin{eqnarray}
\Gamma_{NJL}(\sigma, \pi) &=& \int dx_2 dx_3 \int dx_4 dx_1 L_{NJL}^{(eff)}
(\sigma(x), \pi(x)) \nonumber  \\ 
&\simeq& \frac{1}{2\pi} |eB| a^4 \sum 
\limits_{i,j = - \infty}^{\infty} \sum \limits_{n,m = - \infty}^{\infty}  
\tilde{L}_{NJL} ^{(eff)}(\sigma_{ij}(n,m), \pi_{ij}(n,m)) 
\label{eq28} 
\end{eqnarray}            
where $\sigma_{ij}(n,m) = \sigma(x)$, $\pi_{ij}(n,m) = \pi(x)$, with
$x_2=ia$, $x_3=ja$, $x_4=ix_0=na$, $x_1=ma$, and here the factor
$|eB|/2\pi$ was explicitly factorized from $L_{NJL}^{(eff)}$.
Now, taking into account Eqs.(\ref{eq7}),(\ref{eq12}) and Eqs.(\ref{eq27}),
(\ref{eq28}), we find that 
\begin{eqnarray}
\Gamma_{NJL} &=& \frac{1}{2\pi} |eB| a^4 \sum \limits_{i,j =
- \infty}^{\infty} \sum \limits_{n,m = - \infty}^{\infty}  \tilde{L}_{NJL}
^{(eff)}(\sigma_{ij}(n,m), \pi_{ij}(n,m)) \nonumber \\
&\to& \sum 
\limits_{x_2, x_3} \int dx_4 dx_1 L_{GN}^{(eff)} (\sigma_{x_2 x_3} 
(x_{\parallel}), \pi_{x_2 x_3} (x_{\parallel})) 
\label{eq29} 
\end{eqnarray} 
in the ``continuum" limit with
$(|eB|/2\pi)a^2\equiv\pi|eB|/2\Lambda^2=
\pi/2C$ (here $\Lambda=\pi/a$ is the ultraviolet cutoff on the
lattice)\footnote{Of course, the ultraviolet cutoff on the lattice is
different from the cutoff in the proper-time regularization used above.
However, since the constant $C=\Lambda^2/|eB|$ is anyway arbitrary here,
we use the same notation for the cutoff on the lattice as in the 
proper-time regularization.}.
The lagrangian density ${\cal L}^{(eff)}_{GN}$ corresponds to
the GN model with the number of colors $\tilde{N}_c=(\pi/2C)N_c$. Note
that the symbol $\sum \limits_{x_2, x_3}$
here is somewhat formal and it just implies that the GN
model occurs at each point in the $x_{2}x_{3}$--plane.

The physical meaning of this reduction of the
NJL model in a magnetic field is rather clear. At weak
coupling, the fermion pairing in a magnetic field takes place
essentially for fermions in the LLL with the momentum $k_1$=0.
The size of the radius of the LLL
orbit is $l$=$|eB|^{-1/2}$ \cite{12}. As the
magnetic field goes to infinity,
this
radius shrinks to zero. Then, because of the degeneracy in the LLL
\cite{12}, there are $(|eB|/2\pi)a^2=\pi/2C$ states with $k_1$=0
at each point in the $x_{2}x_{3}$--plane. This degeneracy factor
(proportional to the (local) magnetic flux across a plaquette)
leads to changing the number of colors, 
$N_c\to\tilde{N}_c=(\pi/2C)N_c$, in the GN model.Note that since
$\tilde{N}_c$ appears analytically in the path integral of the theory,
one can give a non-perturbative meaning to the theory with
non-integral $\tilde{N}_c$.

A few comments are in order.

Since these GN models are independent, the parameters of the chiral
$U_{L}(1)\times U_{R}(1)$ transformations can depend on ${\bf x}_\perp$.
In other words, here the chiral group
is $\prod \limits_{x_{\perp}} U^{(x_\perp)}_{L}(1)\times
U^{(x_\perp)}_{R}(1)$. As a result, there are an infinite number
of gapless modes $\pi_{x_{2}x_{3}}(x_{\parallel})$ in the ``continuum"
limit.

Since there is no spontaneous breakdown of continuous symmetries
at $D=1+1$, the fields $\pi_{x_{2}x_{3}}(x_{\parallel})$ do not describe
NG bosons (though they do describe gapless BKT excitations) \cite{8}.

Since the magnetic field depends on $\Lambda$ in the ``continuum" 
limit, it can be considered as an additional parameter
(``coupling constant") in the renormalization group.
The ratio $b=|eB|/\Lambda^2=C^{-1}$ is arbitrary here.
>From the point of view of the renormalization group,
this can be interpreted as the presence of a line of 
ultraviolet fixed points for the dimensionless coupling $b$.
The values of $b$ on the line define the local magnetic flux
and, therefore, the number of colors $\tilde{N}_c$ in the
corresponding GN models. 

Our consideration of reducing the NJL model in a magnetic field to
a continuum set of the GN models was somewhat heuristic. It would be
worth deriving this reduction in a more rigorous way, putting
the NJL model on a lattice in Euclidean space and then realizing 
renormalizations in the ``continuum" limit.

In the next section, we shall discuss the connection between the
GN model and the NJL model in a magnetic field 
in more detail.

\section{More about the Connection between the GN Model 
and the NJL Model in a Magnetic Field}

In the previous section, 
the reduction of
the NJL model in a magnetic field of the infinite strength
to a continuum set of the
GN models
was established. But what is the connection
between the NJL and GN models at finite,
though large, values of the magnetic
field?

In order to answer this question, let us turn to a more detailed
discussion of the infrared dynamics within these two models.

The GN model is asymptotically free, with the bare coupling constant
$\tilde{G}=2\pi/\tilde{N}_c\ln(\Lambda^2/m_{dyn}^2)\to 0$ as
$\Lambda\to\infty$. Therefore there is dimensional transmutation
in the model: in the scaling region ($\tilde{G}\ll 1$) the
infrared dynamics, with momenta $k$ satisfying
$[\ln(\Lambda^2/k^2)]^{-1}\ll 1$, 
are essentially independent of either the coupling constant
$\tilde{G}$ or the cutoff $\Lambda$; the only relevant parameter is
the dynamical mass $m_{dyn}$ (which is an analogue of the parameter
$\Lambda_{QCD}$ in QCD). Because of Eq.(\ref{eq21}) for $m_{dyn}$, one 
might expect that a similar dimensional transmutation should take
place in the NJL model in a magnetic field: for $|eB|GN_c\ll 1$, the
infrared dynamics, with $k^2\ll |eB|$, should be essentially
independent of the magnetic field and the coupling $G$.

The real situation, however, is  more subtle. Let us look at the
propagators for fermions and $\sigma$ and $\pi$ particles  at low momenta 
in the NJL model with a magnetic field,
in leading order in $1/N_c$ \cite{4}:
\begin{eqnarray}
\tilde{S}^{(0)} (k)&=&i\exp\left(-\frac{{\bf k}^2_{\perp}}{|eB|}\right)
         \frac{k_0\gamma^0-k_1\gamma^1+m_{dyn}}{k_0^2-k_1^2-m_{dyn}^2}
         \left(1-i\gamma^2\gamma^3\mbox{sign}(eB)\right),\label{eq30}\\
{\cal D}_{\sigma}(k)&=&\frac{C_{\sigma}}{N_c}\Bigg[k_0^2-k_1^2-
        \frac{3m_{dyn}^2}{|eB|}
        \ln\left(\frac{|eB|}{\pi m_{dyn}^2}\right){\bf k}^2_{\perp}
        -12m_{dyn}^2\Bigg]^{-1},\label{eq31}\\
{\cal D}_{\pi}(k)&=&\frac{C_{\pi}}{N_c}\Bigg[k_0^2-k_1^2-
        \frac{m_{dyn}^2}{|eB|}
        \ln\left(\frac{|eB|}{\pi m_{dyn}^2}\right){\bf k}^2_{\perp}
        \Bigg]^{-1}, \label{eq32}
\end{eqnarray}
where ${\bf k}_{\perp}=(k_2,k_3)$ and where $C_{\sigma}$ and $C_{\pi}$ are
some inessential constants (we consider the case of the weak
coupling, when relation (\ref{eq21}) is valid). Because of relations 
(\ref{eq21}) and (\ref{eq23}), the coefficients of the ${\bf
k}^2_{\perp}$--terms in the propagators ${\cal D}_{\sigma}(k)$ and
${\cal D}_{\pi}(k)$ are exponentially small. Also, in the  infrared
region, the fermion propagator is independent (up to power
corrections $\sim({\bf k}^2_{\perp}/|eB|)^n$, $n\geq 1$) of the
magnetic field. Therefore one might think that in this case, like in the
GN model, the only relevant parameter for the infrared dynamics is
$m_{dyn}$. However, while the dependence of the propagators of
fermions and $\sigma$ particles on $|eB|$ can indeed be neglected,
this dependence
is essential in the case of the propagator of the gapless NG mode
$\pi$. The ${\bf k}^2_{\perp}$--term in ${\cal
D}_{\pi}(k)$ provides the ($3+1$)--dimensional character of the NG
propagator which, as explained in Introduction, is
crucial for the realization of spontaneous chiral symmetry breaking
in the model. In a sense, the magnetic length $l=|eB|^{-1/2}$ plays
here the same role as the $\epsilon$--parameter in the
($2+\epsilon$)--dimensional GN model \cite{5}.

As $|eB|\to\infty$, the NJL model is reduced to a set of 
independent GN models. However, at finite values of $|eB|$, the 
transverse velocity of the NG mode, though small 
[$|{\bf v}_{\perp}| \leq \sqrt{(m_{\rm dyn}^2/|eB|)
\ln(|eB|/\pi m_{\rm dyn}^2)}$], is not zero. Therefore there are 
now transitions of field quanta between different points in 
$x_{2}x_{3}$--plane, i.e.  interactions occur between the GN 
models associated with different points $\bf x_\perp$.  As a 
result, at finite $|eB|$ the chiral group is reduced to global 
$U_{L}(1)\times U_{R}(1)$ transformations, and there is only one 
NG boson $\pi$ for this case.

Therefore the reduction of the NJL model, described in the
previous section, takes place only as $|eB|\to\infty$. At finite
values of the magnetic field, the dynamics in the NJL and GN
models are different: while there is spontaneous chiral symmetry
breaking in the NJL model, the BKT phase is realized in the
GN model \cite{8}. The connection between these two sets of dynamics
is similar to that between the dynamics of 2--dimensional
and ($2+\epsilon$)--dimensional GN models.

In conclusion, we emphasize that this discussion pertains only to the
NJL model with a weak coupling constant, when relation (\ref{eq21})
is valid. In the case of the NJL model with a near--critical $g$, the
situation is different: when $g\to g_c-0$, (\ref{eq22}) is valid. 
The difference between these two dynamical regimes reflects the
fact that, while at weak coupling the LLL dominates, at
near--critical $g$ all Landau levels are relevant \cite{4}.

In the next section, we shall discuss the dynamics in the
supersymmetric NJL model.

\section{Chiral Symmetry Breaking in the Supersymmetric 
NJL Model in a Magnetic Field}

As is well known, there is no spontaneous chiral symmetry breaking 
in the supersymmetric (SUSY) NJL model at any value of the coupling 
constant \cite{6}. Here we shall show that an external magnetic field
changes the situation dramatically: chiral symmetry breaking occurs
for all (positive) values of the coupling constant. Moreover, at weak
coupling, the dynamics in SUSY and ordinary NJL models are similar
(and, therefore, intimately connected with
the dynamics in the ($1+1$)--dimensional
GN model).

The action of the SUSY NJL model with the $U_L(1)\times U_R(1)$
chiral symmetry in a magnetic field is
\begin{eqnarray}
\Gamma&=&\int d^8z 
    \left[ \bar{Q}e^{V}Q+\bar{Q}^ce^{-V}Q^c 
       + G(\bar{Q}^c\bar{Q})(QQ^c)\right] .
\label{eq33}
\end{eqnarray}
Here we utilize the notations of Ref.\cite{13}, except for our 
choice of metric $g^{\mu\nu}=\mbox{diag}(1,-1,-1,-1)$. In Eq.(\ref{eq33}),
$d^8z=d^4xd^2\theta d^2\bar{\theta}$, $Q^{\alpha}$ and $Q^c_{\alpha}$ 
are chiral superfields carrying the color index 
$\alpha=1, 2, \dots, N_c$, 
i.e. 
$Q^{\alpha}$ and $Q^c_{\alpha}$
are assigned to the fundamental and antifundamental representations of
the $SU(N_c)$, respectively:
\begin{eqnarray}
          Q^{\alpha}  = \varphi^{\alpha} 
                      + \sqrt{2}\theta\psi^{\alpha}
                      + \theta^2F^{\alpha} , &\quad&
         Q^c_{\alpha} = \varphi^c_{\alpha}
                      + \sqrt{2}\theta\psi^c_{\alpha}
                      + \theta^2F^c_{\alpha} 
\label{eq34}
\end{eqnarray}
(henceforth we will omit color indices). The vector superfield 
$V(x,\theta,\bar{\theta})
=-\theta\sigma^{\mu}\bar{\theta}A^{ext}_\mu$,
with 
$A^{ext}_\mu = B x^2 \delta_{\mu}^3$, 
describes an external magnetic field which is in the $+x_1$ direction.

The action (\ref{eq33}) is equivalent to the following action:
\begin{eqnarray}
\Gamma_{A} &=& \int d^8z 
           \left[\bar{Q}e^{V}Q+\bar{Q}^ce^{-V}Q^c
                 + \frac{1}{G}\bar{H}H\right]-
           \nonumber\\
               &&- \int d^6z\left[\frac{1}{G}HS-QQ^cS\right]
                 - \int d^6\bar{z}\left[\frac{1}{G}\bar{H}\bar{S}
                 -                \bar{Q}\bar{Q}^c\bar{S}\right] .
\label{eq35}
\end{eqnarray}
Here 
$d^6z=d^4xd^2\theta$, 
$d^6\bar{z}=d^4xd^2\bar{\theta}$,
and $H$ and $S$ are two auxiliary chiral fields:
\begin{eqnarray}
      H=h+\sqrt{2}\theta\chi_h+\theta^2f_h , &\quad&
      S=s+\sqrt{2}\theta\chi_s+\theta^2f_s .
\label{eq36}
\end{eqnarray}
The Euler--Lagrange equations for these auxiliary fields take the
form of constraints:
\begin{eqnarray}
               H = GQQ^c, &\quad&
               S = - \frac{1}{4}\bar{D}^2(\bar{H})=
                   - \frac{G}{4}\bar{D}^2(\bar{Q}\bar{Q}^c).
\label{eq37}
\end{eqnarray}
Here $\bar{D}$ is a SUSY covariant derivative \cite{13}. The action 
(\ref{eq35}) reproduces Eq.(\ref{eq33}) upon application of the
constraints (\ref{eq37}).

In terms of the component fields, the action (\ref{eq35}) is
\begin{eqnarray}
\Gamma_{A} &=& \int d^4x \Bigg[
- \varphi^{\dagger}(\partial_{\mu}-ieA^{ext}_{\mu})^2\varphi
- \varphi^{c\dagger}(\partial_{\mu}+ieA^{ext}_{\mu})^2\varphi^c        
\nonumber \\
&&+ i\bar{\psi}\bar{\sigma}^{\mu}(\partial_{\mu}-ieA^{ext}_{\mu})\psi
+ i\bar{\psi}^c\bar{\sigma}^{\mu}(\partial_{\mu}+ieA^{ext}_{\mu})\psi^c
\nonumber \\
&&+ F^{\dagger}F + F^{c\dagger}F^c 
+ \frac{1}{G}\left( -h^{\dagger}\Box h 
              + i\bar{\chi}_h\bar{\sigma}^{\mu}\partial_{\mu}\chi_h 
              + f^{\dagger}_hf_h \right)
\nonumber \\
&&+ \frac{1}{G}\left( \chi_h\chi_s - hf_s- sf_h + h.c.\right)
\nonumber \\
&&- \Big( s\psi\psi^c + (\varphi\psi^c+\varphi^c\psi)\chi_s
         - s(\varphi F^c + \varphi^cF) - \varphi\varphi^c f_s 
         + h.c.\Big)
         \Bigg] .
\label{eq38}
\end{eqnarray}
Let us consider the effective potential in this model. For this
purpose, one can treat all the auxiliary scalar fields as (independent of $x$)
constants and all the auxiliary fermion fields equal zero (since the
auxiliary fields are colorless, loop diagrams involving them do not contribute
to the effective potential in leading order in $1/N_c$). Then, using
the Euler--Lagrange equations for the fields $F$, $F^c$, $f_h$, $h$
and their conjugates, we find $F^{\dagger}=-s\varphi^c$,
$F^{c\dagger}=-s\varphi$,  $f^{\dagger}_h=s$, $f^{\dagger}_s=0$,
plus h.c. equations. Then the action becomes
\begin{eqnarray}
\Gamma_{A} &=& \int d^4x \Bigg[
- \varphi^{\dagger}\left[(\partial_{\mu}-ieA^{ext}_{\mu})^2
                    + \rho^2 \right]\varphi
- \varphi^{c\dagger}\left[(\partial_{\mu}+ieA^{ext}_{\mu})^2
                    + \rho^2 \right] \varphi^c        
\nonumber \\
&&+ i\bar{\psi}_D\gamma^{\mu}(\partial_{\mu}
        -ieA^{ext}_{\mu})\psi_D - \sigma\bar{\psi}_D\psi_D
        - \pi\bar{\psi}_Di\gamma^5\psi_D - \frac{\rho^2}{G}
\Bigg] ,
\label{eq39}
\end{eqnarray}
where $s=\sigma+i\pi$, $\rho^2=|s|^2=\sigma^2+\pi^2$, and the Dirac
fermion field $\psi_D$ is introduced.

In leading order in $1/N_c$, 
the effective potential $V(\rho)$ can now be derived in the
same way as in the ordinary NJL model \cite{4}. The difference is that,
besides fermions, the two scalar fields $\varphi^c$ and $\varphi$
give a contribution to $V(\rho)$:
\begin{equation}
V(\rho) = \frac{\rho^2}{G} 
               + V_{fer}(\rho) + 2 V_{bos}(\rho),
\label{eq40}
\end{equation}
where 
\begin{eqnarray}
V_{fer}(\rho) &=& \frac{N_c}{8\pi^2 l^4} 
               \int\limits^\infty_{1/(l\Lambda)^2} 
               \frac{ds}{s^2}\exp\left(-s( l\rho)^2\right) \coth{s},
\label{eq41} \\
V_{bos}(\rho) &=& -\frac{N_c}{16\pi^2 l^4} 
               \int\limits^\infty_{1/(l\Lambda)^2} 
               \frac{ds}{s^2}\exp\left(-s( l\rho)^2\right) 
               \frac{1}{\sinh{s}} .
\label{eq42}
\end{eqnarray}
As is shown in the Appendix, the potential $V(\rho)$ can be
rewritten as
\begin{eqnarray}
V(\rho) &=& \frac{N_c}{8\pi^2 l^4} 
        \Bigg[ \frac{( l\rho)^2}{g}
           + ( l\rho)^2\left(1-\ln\frac{( l\rho)^2}{2}\right)
           + 4\cdot\int\limits_{( l\rho)^2/2}^{[( l\rho)^2+1]/2}
             dx \ln\Gamma(x)\Bigg]+
        \nonumber\\
          &+& \frac{N_c}{16\pi^2 l^4} 
        \Bigg[ \ln(\Lambda l)^2-\gamma -\ln(8\pi^2) \Bigg]
            + O\left(\frac{1}{\Lambda}\right),
\label{eq43}
\end{eqnarray}
where the dimensionless coupling constant is $g=GN_c/8\pi^2l^2$.

As the magnetic field $B$ goes to zero ($l\to\infty$), we recover the
known expression for the potential in the SUSY NJL model \cite{6}:
\begin{equation}
V(\rho) = \frac{\rho^2}{G} .
\label{eq44}
\end{equation}
The potential
$V(\rho)$ is positive--definite, as has to be in a 
supersymmetric theory. The only minimum of this potential
is $\rho=0$ corresponding to the chiral symmetric vacuum.

The presence of a magnetic field changes this situation
dramatically: at $B\neq 0$, a non--trivial global minimum,
corresponding to spontaneous chiral symmetry breaking, exists for all
$g>0$. Moreover, we show below that, as the coupling $g\to 0$, the
SUSY NJL model becomes equivalent to the ordinary NJL model and,
therefore, at weak coupling and as $B\to\infty$, it is reduced to
the same continuum set of the $(1+1)$--dimensional GN models.
On the other hand, the dynamics of the SUSY and non--SUSY NJL models in 
a magnetic field are very different at strong coupling.

The gap equation $dV/d\rho=0$, following from Eq.(\ref{eq43}), is
\begin{equation}
\frac{N\rho}{4\pi^2 l^2}
      \Bigg[ \frac{1}{g}-\ln\frac{( l\rho)^2}{2}
          + 2\ln\Gamma\left(\frac{( l\rho)^2+1}{2}\right)
          - 2\ln\Gamma\left(\frac{( l\rho)^2}{2}\right)
      \Bigg] = 0.
\label{eq45}
\end{equation}
As can be seen from (\ref{eq43}), at $B\neq 0$ the trivial
solution $\rho=0$ to this equation corresponds to a maximum of $V$: 
$d^2V/d\rho^2|_{\rho=0}=-\infty$. Numerical analysis of equation 
(\ref{eq45}) for $g>0$ and $B\neq 0$ shows that there is a nontrivial 
solution $\rho=\bar{\sigma}=m_{dyn}$ which is the
global minimum of the potential. The analytic expression for $m_{dyn}$
can be obtained at small $g$ (when $m_{dyn}l\ll 1$) and very large
$g$ (when $m_{dyn}l\gg 1$).
In those two cases, the results are: 

{\it a)} $g\ll 1$ ($m_{dyn}l\ll 1$). 
The gap equation (\ref{eq45}) is 
\begin{equation}
            \frac{1}{g}\simeq -\ln \frac{\pi(\rho l)^2}{2} ,
\label{eq46}
\end{equation}
i.e.
\begin{equation}
m_{dyn}\simeq \sqrt{\frac{2|eB|}{\pi}}
              \exp\left[-\frac{1}{2g}\right]
              = \sqrt{\frac{2|eB|}{\pi}}
              \exp\left[-\frac{4\pi^2}{|eB|N_cG}\right] .
\label{eq47}
\end{equation}

{\it b)} $g\gg 1$ ($m_{dyn}l\gg 1$). Now, the gap 
equation (\ref{eq45}) gives
\begin{equation}
            \frac{1}{g}\simeq \frac{1}{2(\rho l)^2} ,       
\label{eq48}
\end{equation}
i.e.
\begin{equation}
m_{dyn}\simeq\sqrt{\frac{g|eB|}{2}}
       = \frac{|eB|}{4\pi}\sqrt{GN_c} .
\label{eq49}
\end{equation}
The numerical solution to the gap equation (\ref{eq45}) for general
values of $g$ is shown in Fig.~1.

Let us discuss the case of the weak coupling in more detail. By using
the asymptotic series for $\Gamma(x)$ \cite{10}, the potential
(\ref{eq43}) can be rewritten at $\rho l \ll 1$ as
\begin{eqnarray}
V(\rho) &=& V(0) + \frac{N_c}{8\pi^2l^4} 
    \Bigg[ \frac{(\rho l)^2}{g} + (\rho l)^2 
           \left( \ln\frac{(\rho l)^2}{2} - 1 \right)+
    \nonumber\\
        && \qquad + (\rho l)^2 \ln\pi + O((\rho l)^4) \Bigg] .
\label{eq50}
\end{eqnarray}
Using Eq.(\ref{eq46}), the coupling constant $g$ can be expressed
through $m_{dyn}$, and the potential can be rewritten as
\begin{eqnarray}
V(\rho) &=& V(0) + \frac{N_c\rho^2}{8\pi^2l^2} 
    \Bigg[ \ln\frac{\rho^2}{m_{dyn}^2} - 1 + O((\rho l)^2) \Bigg] .
\label{eq51}
\end{eqnarray}
This potential coincides (up to exponentially small terms) 
with that in the ordinary (weakly coupling)
NJL model (see Eq.(\ref{eq26})). 

Now, let us consider the kinetic term in the SUSY NJL model in a 
magnetic field.  As shown in the Appendix, at weak coupling, 
this term also coincides with that in the ordinary NJL model. 
Therefore, at weak coupling, these two models are equivalent. 

Note that the equivalence between the SUSY and non--SUSY NJL 
models becomes explicit only after the renormalization of the 
coupling constants.  The structure of the ultraviolet 
divergences in the effective potentials of these models is quite 
different (compare Eqs.  (\ref{eq13}) and (\ref{eq43})). This 
reflects the point that only the infrared (and not ultraviolet) 
dynamics in these models are equivalent, i.e., in the 
renormalization group language, these two models are assigned to 
the same universality class \cite{14}.

The physical picture underlying this equivalence is clear. An 
external magnetic field explicitly breaks supersymmetry: the 
spectra of charged free fermions and bosons in a magnetic field 
are essentially different \cite{12}. While for fermions the 
spectrum is 
\begin{equation} E_n(k_1)=\pm\sqrt{m^2+2|eB|n+k_1^2} , 
\qquad n=0,1,2,\dots  , \label{eq52} 
\end{equation} 
for bosons it is 
\begin{equation} 
E_n(k_1)=\pm\sqrt{m^2+|eB|(2n+1)+k_1^2} , 
\qquad n=0,1,2,\dots .  \label{eq53} 
\end{equation} 
The crucial difference between them is that while for fermions 
the energy of the LLL is independent of $B$ and (at $k_1=0$) 
goes to zero as $m\to 0$, for bosons the energy of the LLL is 
$E_0(k_1)=\pm\sqrt{|eB|+k_1^2}$. In other words, there is a gap 
$\Delta E = \sqrt{|eB|}$ in the spectrum of massless bosons in a 
magnetic field. Recall that, at weak coupling, the chiral 
condensate in the NJL model occurs because of the fermion 
pairing  in the LLL with $k_1=0$. This also happens in the SUSY 
NJL model, in which the bosonic degrees of freedom become 
irrelevant at weak coupling.

Notice that, in the effective action, the field $\rho$ plays the role
of the mass. In particular, the infrared singularities (as $\rho\to
0$) in the expressions for the potential and the 
kinetic term (see Eqs.(\ref{eq26}) and
(\ref{eq27})) reflect the absence of a gap in the
LLL for massless fermions.

We emphasize that the equivalence between these two models takes 
place at weak coupling only. When $g$ becomes larger than 
$g_c=1$, the dynamics in these models are essentially different.  
While at $g>g_c=1$ in the ordinary NJL model, the chiral 
symmetry is spontaneously broken even without an external 
magnetic field, there is no spontaneous chiral symmetry breaking 
in the SUSY NJL model as $B\to 0$ at any value of $g$ (see 
Eq.(\ref{eq49})). The difference between the dynamical regimes 
with weak and strong coupling corresponds to the fact that, 
while at weak $g$ the dynamics of the LLL of fermions dominates, 
at $g\geq g_c=1$ all (fermionic and bosonic) Landau levels 
become relevant.

\section{Conclusion.}

In this paper we have studied the infrared dynamics of both 
ordinary and SUSY NJL models in a magnetic field. It has been 
shown that, at weak coupling, the infrared dynamics in these two 
models are equivalent and, as $|eB|\to\infty$, the models reduce 
to a continuum set of $(1+1)$--dimensional GN models.

In this paper, as in Refs.\cite{1,2,3,4}, only the case of a 
homogeneous magnetic field has been considered. To extend the 
present results to inhomogeneous field configurations, we note 
first that the number of colors in the GN model is 
$\tilde{N}_c=(\pi/2C)N_c$, where the factor $\pi/2C= 
a^2|eB|/2\pi$ is proportional to the $local$ magnetic flux 
attached to each point in $x_{2}x_{3}$--plane. Let us consider a 
magnetic field ${\bf B}({\bf x}_\perp)$, directed in $+x_1$ 
direction but depending on the transverse coordinates. It is 
tempting to speculate that in this case, as $|eB|({\bf 
x}_{\perp})\to\infty$, the NJL model will be reduced to a set of 
the GN models with the number of colors $\tilde{N}_c({\bf 
x}_{\perp})=(\pi/2C({\bf x}_{\perp}))N_c$ (where $C({\bf 
x}_{\perp})=\Lambda^2/|eB|({\bf x}_{\perp})$), which is 
different in different points of the $x_{2}x_{3}$-plane. It 
would be interesting to check these speculations by studying the 
NJL model in inhomogeneous field configurations.

The results of the present paper are in agreement with the 
general conclusion of Refs.\cite{1,2,3,4}, that the catalysis of 
chiral symmetry breaking by a magnetic field is a universal, 
model independent effect. This catalysis may have possible 
applications to cosmology, particle physics and condensed matter 
physics \cite{4,15}.

A specific application of interest would be a linkage between 
catalysis of chiral symmetry breaking and the existence of very 
strong primordial magnetic fields in the early universe 
\cite{16}. The results of the present paper may be especially 
relevant for cosmological models based on supersymmetric 
dynamics \cite{9}.

Also, the effect of the dimensional reduction by external fields 
may be quite general and relevant for multi-dimensional field 
theories. 

\section*{Acknowledgments}

This research is supported in part by the Natural Science and
Engineering Research Council of Canada (NSERC).

\appendix
\section*{Effective Action in the SUSY NJL Model}

In this Appendix, the effective action for the SUSY NJL model in a
magnetic field is derived.

Besides the expression for the fermion propagator in a magnetic
field, which was used in Refs.\cite{2,4}, we also need to know 
the expression for the propagator in a magnetic field for a charged
scalar with the mass $m=\rho$ \cite{17}:
\begin{eqnarray}
D(x,y)=\exp\biggl[{ie\over 2}(x-y)^\mu A_\mu^{\rm ext}(x+y)\biggr]
\tilde D(x-y)~, 
\label{eqA1}
\end{eqnarray}
where the Fourier transform of $\tilde D(x)$ is
\begin{eqnarray}
\tilde{D}(k) &=& - \int^\infty_0 \frac{ds}{\cosh{eBs}}
             \exp \Bigg[-s\left( \rho^2-k_0^2+{\bf k}_{\perp}^2
             \frac{\tanh (eBs)}{eBs}+k_3^2 \right)\Bigg] .
\label{eqA2}
\end{eqnarray}
The expression (\ref{eq40}) for the effective potential in the 
SUSY NJL
model can be obtained in the same way as for the ordinary NJL model
in Ref.\cite{4}. Let us show that this expression is equivalent to
expression (\ref{eq43}).

Eq.(\ref{eq40}) can be rewritten as 
\begin{eqnarray}
V(\rho)&=&\frac{\rho^2}{2G}+\frac{N_c}{16\pi^2l^4} 
       \Bigg[\ln\left(\frac{\Lambda}{\rho}\right)^2
     - \gamma + 2 I((\rho l)^2)\Bigg]
     + O\left(\frac{1}{\Lambda}\right)
\label{eqA3}
\end{eqnarray}
where 
\begin{eqnarray}
I(\beta)=\int\limits_0^{\infty} dse^{-\beta s}
        \left[\frac{\cosh s -1 }{s^2\sinh s}-\frac{1}{2s}\right],
        \qquad \beta>0.
\label{eqA4}
\end{eqnarray}
Let us show that
\begin{eqnarray}
I(\beta)=\beta\left(1-\ln\frac{\beta}{2}\right)
        + \frac{1}{2}\ln{\beta} - \frac{1}{2}\ln{8\pi^2}
        + 4\int\limits_{\beta/2}^{(\beta+1)/2} dx \ln\Gamma(x).
\label{eqA5}
\end{eqnarray}
Using the integral representation for generalized zeta function
\cite{10}, we find
\begin{eqnarray}
I(\beta)&=&\lim_{\mu\to-1}I(\beta,\mu)\equiv
   \lim_{\mu\to-1}\int\limits_0^{\infty} ds s^{\mu}e^{-\beta s}
        \left[\frac{\cosh s -1 }{s\sinh s}-\frac{1}{2}\right]=
        \nonumber\\
        &=&\lim_{\mu\to-1}\Gamma(\mu)\Bigg[ 
        2^{1-\mu}\zeta\left(\mu,\frac{\beta}{2}\right)-\beta^{-\mu}
        -2^{1-\mu}\zeta\left(\mu,\frac{\beta+1}{2}\right)
        - \frac{\mu}{2} \beta^{-1-\beta}\Bigg].
\label{eqA6}
\end{eqnarray}
In order to find  this limit, we need the following identity:
\begin{eqnarray}
\left.\frac{\partial}{\partial z} 
\Bigg[\zeta(z,q_2)-\zeta(z,q_1)\Bigg]
      \right|_{z=-1}=\frac{q_2-q_1}{2}(q_2 + q_1 - 1 - \ln{2\pi})
     + \int\limits_{q_1}^{q_2} dq \ln\Gamma(q).
\label{eqA7}
\end{eqnarray}
To derive it, we use the relation \cite{10}
\begin{eqnarray}
\frac{\partial}{\partial q} \zeta(z,q)=-z\zeta(z+1,q).
\label{eqA8}
\end{eqnarray}
Differentiating it with respect to $z$ and integrating over $q$, we
obtain
\begin{eqnarray}
\frac{\partial}{\partial z} \Bigg[\zeta(z,q_2)-\zeta(z,q_1)\Bigg]
      =-\int\limits_{q_1}^{q_2} d q \zeta(z+1,q)
      - z \int\limits_{q_1}^{q_2} dq 
        \frac{\partial}{\partial z} \zeta(z+1,q).
\label{eqA9}
\end{eqnarray}
Then, using the identities \cite{10}
\begin{eqnarray}
\zeta(0,q)=\frac{1}{2}-q, \qquad 
\left.\frac{\partial}{\partial z} \zeta(z,q) \right|_{z=0}
      =\ln\Gamma(q)-\frac{1}{2}\ln{2\pi},
\label{eqA10}
\end{eqnarray}
we obtain Eq.(\ref{eqA7}).

>From Eqs.(\ref{eqA6}), (\ref{eqA7}) and relations \cite{10}
\begin{eqnarray}
\Gamma(\mu)=\frac{\Gamma(\mu+2)}{(\mu+1)\mu}, \qquad 
\zeta(-1,q)=-\frac{1}{2}\left(q^2-q+\frac{1}{6}\right) ,
\label{eqA11}
\end{eqnarray}
we obtain equation (\ref{eqA5}).

Eqs.(\ref{eqA3}) and (\ref{eqA5}) lead to relation  (\ref{eq43}) for
the effective potential.

Let us show that, at weak coupling, the kinetic term in the effective
action of the SUSY NJL model coincides with that in the ordinary NJL
model. For this purpose, we shall prove that the contribution of the
scalar fields in the kinetic term is suppressed.

By using the approach of Ref.\cite{4}, one finds that the functions
$f_1^{\mu\nu}$ and $f_2^{\mu\nu}$ in the kinetic term (\ref{eq14})
are now equal to:
\begin{eqnarray}
f_1^{\mu\nu}&=&\left(f_1^{\mu\nu}\right)_{fer},
\label{eqA12}\\
f_2^{\mu\nu}&=&\left(f_2^{\mu\nu}\right)_{fer}-2i\rho^2
   \int\frac{d^4k}{(2\pi)^4}\tilde{D}(k)
\frac{\partial^2\tilde{D}(k)}{\partial k_{\mu}\partial k_{\nu}},
\label{eqA13}
\end{eqnarray}
where $\left(f_i^{\mu\nu}\right)_{fer}$ are as given in (\ref{eq15}).
The crucial point for us is that, while the fermionic contribution
(\ref{eq27}) in $f_{i}^{\mu\nu}$ is divergent as $\rho l\to 0$,
the contribution of the scalars is finite in this limit. This
conclusion, following directly from Eqs.(\ref{eqA2}) and  
(\ref{eqA13}),
reflects the fact that there is a gap in the spectrum of massless
charged scalars in a magnetic field (see Eq.(\ref{eq53})). Since, 
at weak coupling, the parameter  $\rho l$ is exponentially small, 
we  conclude that the contribution of the scalars is indeed 
suppressed at small $g$.


\vspace{1cm}

\centerline{FIGURE CAPTIONS}

Figure 1. The curve of $m_{dyn}l=m_{dyn}/\sqrt{|eB|}$ as the 
function of the inverse coupling constant $1/g$ in the SUSY 
NJL model.

\newpage
\begin{figure}
\epsfxsize=12cm
\epsffile[70 00 470 700]{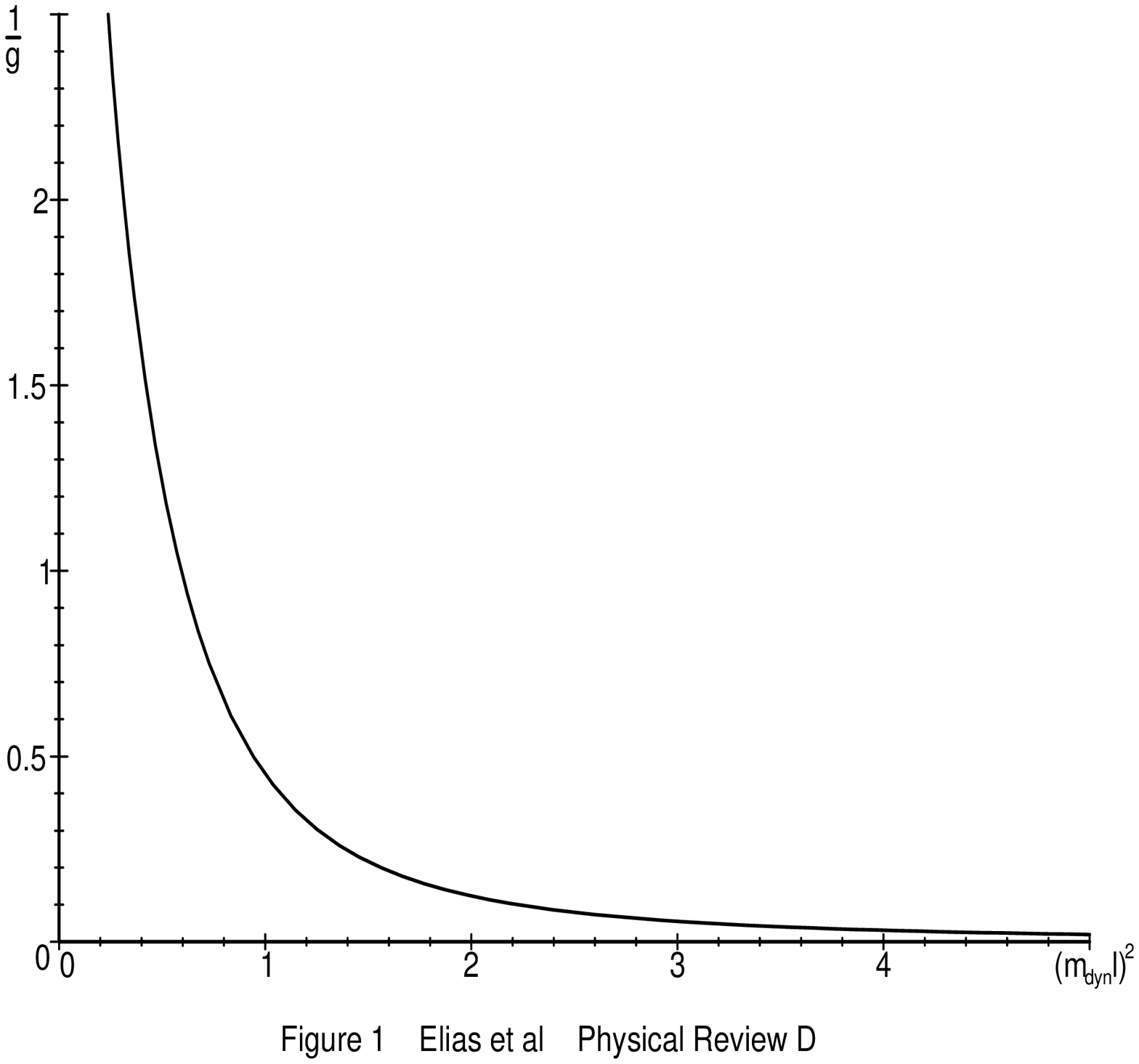}
\end{figure}

\end{document}